# Fluorine effect in layered oxypnictide LaOFeAs

*A. Reyes-Serrato\*, D.H. Galván and G. Soto*
*Universidad Nacional Autónoma de México, Centro de Nanociencias y Nanotecnología*
*Km 107 Carretera Tijuana-Ensenada, Ensenada, Baja California, 22800 México*
*\*armando@cnyn.unam.mx*

## Abstract

Theoretical calculations under the scheme of WIEN2k computer package had been performed on the pristine compound LaOFeAs, as well as on $LaO_{1-x}F_xFeAs$. One factor crucial to the manifestation of superconductivity seems to be fluorine doping; hence we search for its effect. Paying close attention to Fe d-orbitals upon doping of F by O, at first, Fe d-orbitals have $d_z^2$ symmetry, changing gradually to something akin to $d_{x^2-y^2}$ which seems more prone to conductivity than the former one. This issue agree with the accepted results happened in the copper-based high-temperature superconductors.

A new class of high-temperature superconductors discovered earlier this year seems to further challenge BCS theory, that is, the family of doped quaternary layered oxypnictides LnOTmPn (Ln: La, Pr, Ce, Sm; Tm: Mn, Fe, Co, Ni; Pn: P, As). The new superconductors behave in a different way than the cuprate superconductors do, and are thus cataloged as unconventional superconductors (1). The oxypnictides materials are formed by alternating stacks of LnO and TmPn layers (2). In these materials there is one factor that appears to be crucial to the mechanisms and manifestations of superconductivity, that is, the fluorine doping of LnO layers. The change of some oxygen atoms by fluorine induces superconductivity with record high of 52 K in $NdFeAsO_{0.89}F_{0.11}$ and $PrFeAsO_{0.89}F_{0.11}$ (3, 4). At this moment there is not an agreement about the mechanism which originates superconductivity. However if an understanding of the role of fluorine in the electronic structure of oxypnictides is achieved we can shed some light in the mechanisms of superconductivity. With that aim in mind a study of the $LaO_{1-x}F_xFeAs$ compounds with and without fluorine doping by density functional calculations was performed. Especially, paying attention to the behavior of the charge density located in the vicinity of the Fermi level, which is assumed to be responsible for superconductivity.

The theoretical calculations had been performed on the pristine compound LaOFeAs, as well as on $LaO_{1-x}F_xFeAs$. We have employed WIEN2k code (5), that program package allows to perform electronic structure calculations of solids using density functional theory (DFT). It is based on the full-potential (linearized) augmented plane-wave ((L)APW) + local orbitals (lo) method.

The crystalline system of the LaOFeAs compound belongs to the tetragonal *P4/nmm* (129) space group with cell parameters of *a* = 4.0355 Å, *c* = 8.7393 Å. To model the oxygen substitution by fluoride was created a corresponding supercell of 3x3x1 with its symmetries reduce down to the *P-4m2* (115) space group and cell parameters of *a* = 12.1065 Å, *c* = 8.7393 Å. The experimental lattice parameters obtained by Kamihara *et al.*, (2) and z coordinate for La and As, reported by Sing and Du (6) were used.

Our analysis consists in an insight of the electrons with energies very close to the Fermi level. As it can be seen in Figure 1, at this particular energy the electrons are confined to the space around Fe atoms. The computational result shows that the substitution of some oxygen atoms by fluorine provokes an amendment of the electron-filling-order near the Fermi level of Fe *d*-orbitals, in close conformity with the crystalline field theory. The effect of fluoride can be comprehended as a localized perturbation to the crystalline field. Figure 1B yielded information for the unperturbed charge distribution for nonsuperconductive LaOFeAs as dominated by orbital of $d_z^2$ symmetry. However as some oxygen atoms are replaced with fluorine the symmetry inside the perturbed crystalline field changes to something akin to $d_{x^2-y^2}$ orbital in Figure 1C. When the oxygen substitution is enough to reach the $LaO_{0.89}F_{0.11}FeAs$ composition the perturbation is spread out to the entire Fe-layer, as it is show in Figure 1D. In conclusion, the main effect of fluorine in the electronic structure of oxypnictides is to change the orbital symmetry at Fermi level. Why this symmetry is favorable to the manifestation of superconductive? It seems that the $d_{x^2-y^2}$ orbital inside the tetrahedral field allow a better electronic interchange than the $d_z^2$ orbital, but a comprehensive rationalization still has to be elucidated. This insight is an important step toward the understanding of how oxypnictides superconductors work, and it could help researchers to propose new and even better materials.

**Acknowledgments**

Funding support by projects CONACYT 52927 and PAPIIT IN107508

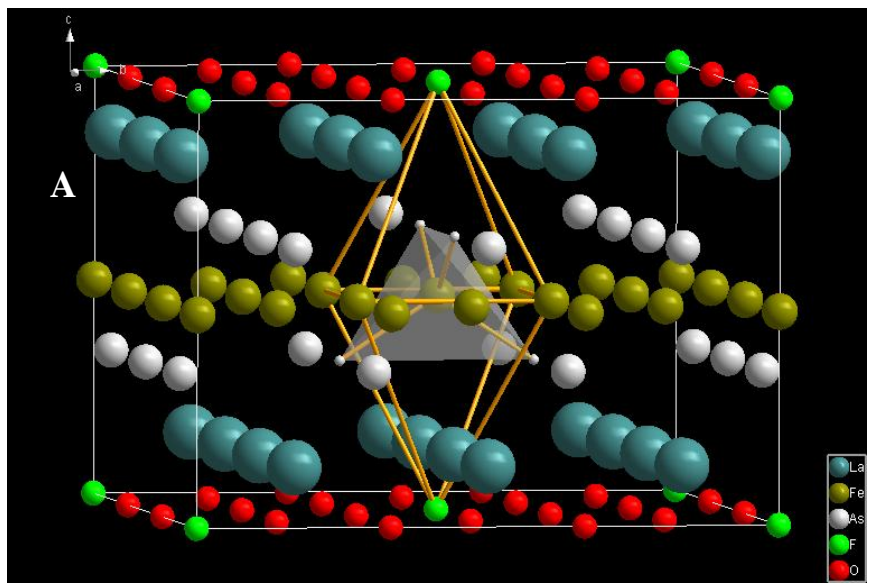
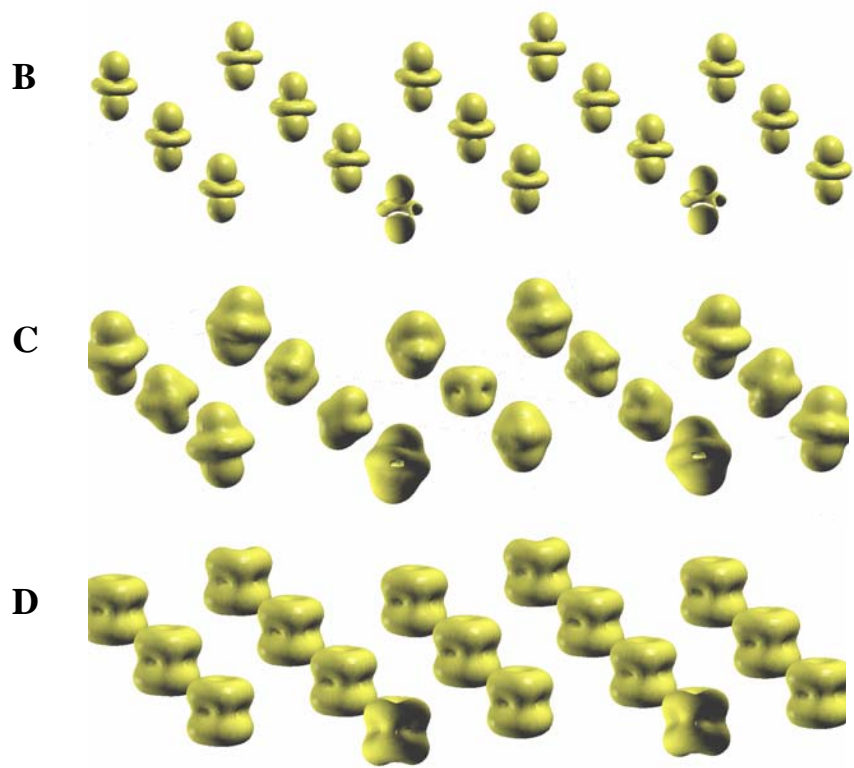

**Fig. 1.** **Distribution analysis of electron with energies close to the Fermi Level.** In **(A)** it is show that the Fe atoms in LaOFeAs are within a tetrahedral crystalline field formed by four As atoms contiguous to them; whilst the As-Fe complex is inside a surrounding octahedral field created by four Fe atoms of the same plane, and two oxygen atoms from the upper and lower planes. **(B)** Shows that the nonsuperconductive state in LaOFeAs is dominated by orbital of $d_z^2$ symmetry. In **(C)** it is show that by replacing some oxygen atoms with fluorine (x = 0.05) the effect is to change the symmetry of $d$-orbitals inside the perturbed crystalline field. When the oxygen substitution is enough to reach the LaO$_{0.89}$F$_{0.11}$FeAs composition the perturbation was extended out to the entire Fe-layer, as it is show in **(D)**.